# Exploring Motion: Integrating Arduino in Physics Education for 21st Century Skills

Atakan Coban[1] & Mert Büyükdede[2]

[1]Dokuz Eylül University, Education Faculty of Buca, Department of Mathematics and Science Education, Physics Education (Chair of Physics Education, Faculty of Physics, Ludwig-Maximilians-Universität München)

[2]Dokuz Eylül University, Education Faculty of Buca, Department of Mathematics and Science Education, Physics Education Division, Buca, Izmir, Turkey.

e-mail: [1]atakancoban39@gmail.com

**Abstract**

In the study, analyzes were made for one-dimensional constant acceleration motion using the Arduino microcontroller and distance sensor, using the position and time values obtained for the movement of an object thrown from bottom to top until it falls to the ground. Within the scope of the study, the concepts of displacement, distance traveled, average speed and their relationships, position-time and speed-time graphs and kinematic equations were analyzed respectively. It was observed that all the results obtained from the analyzes were compatible with the theoretical results obtained using the equations. This study, in which innovative experimental activities and comprehensive analyzes for both constant acceleration motion and motion graphs and flight movements are carried out at a very low cost, has a high reproducibility potential as it contains easily accessible materials. With this type of study, it is very important and necessary to add training for 21st century skills such as material development, programming, and data analysis to physics teaching processes in the age of technology.

## Introduction

Rapid developments in science and technology have brought many changes in the world. The most important aspect of keeping up with these changes is education. Therefore, there is a need to restructure education and training programs depending on these changes. It requires teachers to have different knowledge and skills than they did in the past. For this reason, teachers/candidates should have these skills, which are defined as 21st century skills required by these changes[1].

Skills required for individuals to be aware of the changes occurring in their environment, to adapt to these changes, to follow rapidly developing information and technologies, to reach the right information by analysing the data they have obtained by using high-level thinking skills, and to use this information in their working life and daily life 21st century skills is called[2].

Raising individuals with 21st century skills is possible with an education based on new educational approaches. In this context, the most important educational approach that has come to the fore in recent years is STEM education. STEM education is handled as the organization of science, technology, engineering and mathematics disciplines for the education process. In STEM education, it is important to associate and structure the process with daily life and real life problems[3].

In Physics Curriculums, students not only have knowledge; It is emphasized that they can acquire knowledge and use it to solve the problems they encounter in daily life by transferring the knowledge they have obtained. These skills that individuals use to acquire scientific knowledge are described as scientific process skills[4].

It can be stated that the physics course, which is often recommended to be conducted in a laboratory environment and enriched with experimental activities, allows the acquisition and development of the mentioned scientific process skills[4]. In addition, it is underlined that the Physics course is among the most suitable courses in terms of holistic handling of the disciplines evaluated within the scope of STEM education[5]; The positive effect of using technology together with the teaching process is mentioned[6].

Experimental setups that can be used in fields such as mechanics, optics, electricity, magnetics and thermodynamics can be developed with the Arduino electronic circuit development board, which is the technology component of this study in terms of STEM education[7,8]. Arduino, which is a microcontroller with open-source software and hardware, collects data, records and performs the action by using it with sensors. In addition, it is frequently preferred in the teaching process because it is economical and the programming language is simple[9].

In this study, the Arduino electronic circuit development board was used in an experimental setup in a STEM activity developed for the conceptual understanding of the upward shooting motion. In the study, the upward shooting movement was analysed in detail. Based on these analyses, an inference has been made in a way that will allow the understanding of the concepts in the subject of movement. These concepts are; Displacement, average speed, distance travelled, average speed, motion graphs are mathematical transitions between motion graphs. In summary, it is clearly stated that it is possible to understand all movement issues from the upward shooting movement.

## Theory

In physics, the part that studies the motion of objects, the effects that cause motion, and the equilibrium states of names is called mechanics[10]. It is divided into three parts: mechanics, statics, kinematics and dynamics. In this section, kinematics will be examined. Kinematics is a section of mechanics that studies motion without considering its causes and effects. Kinematics can be grasped by understanding motion and the velocity and acceleration arising from it. Motion can be defined as the continuous displacement of an object from one point to another.

The motion of a particle is completely certain if its position in space is known at every moment. If we know how the position changes according to time, we can find the velocity of the object from here, and how this velocity value changes over time, we can find the acceleration of the object.

Displacement is defined as the difference between the initial ($x_i$) and final ($x_f$) position of a moving body and can be calculated as

$$\overrightarrow{\Delta x} = \overrightarrow{x_f} - \overrightarrow{x_i} \quad (1)$$

The path taken by the moving particle x and the displacement $\overrightarrow{\Delta x}$ of the particle are not the same and should not be confused with each other. For example, let an object start moving from point A and after reaching point B, return to point A again. Here, the path taken by the object is 2AB, but its displacement is zero.

The displacement and the path of moving object are related with average velocity and average speed of it. Velocity is the magnitude of the displacement and speed is the magnitude of the path during in one second. While speed is a scalar quantity without any direction, velocity is a vector quantity and its direction must be specified as well as its magnitude. If displacement is $\overrightarrow{\Delta x}$ and path is x in one dimension, the average velocity $\vec{v}$ and average speed $v$ over a time interval $\Delta t$ are given by

$$\vec{v} = \frac{\overrightarrow{\Delta x}}{\Delta t} \quad (2)$$

$$v = \frac{x}{\Delta t} \quad (3)$$

Changing rate of the velocity in one dimensional motion is equal to acceleration of the movement. Acceleration can be calculated by

$$\vec{a}_{avg} = \frac{\overrightarrow{\Delta v}}{\Delta t} \quad (4)$$

The time-dependent equation of change of position in motion with constant acceleration in one dimension is,

$$x(t) = v_0 t \mp \frac{1}{2} a t^2 \quad (5)$$

where $v_0$ is the initial velocity, a is the acceleration, and t is the elapsed time. Considering that velocity is the change of position per unit time, the first derivative of the equation of change of position with respect to time can be taken to obtain the time-dependent equation of velocity. The equation of variation of velocity obtained by taking the first derivative of Eq. 5 with respect to time is

$$v(t) = \frac{dx(t)}{dt} = v_0 \mp at \quad (6)$$

In some cases, the displacement is known while the elapsed time value is unknown. The timeless velocity equation used to find the final velocity in these cases is

$$v^2(x) = v_0^2 \mp 2ax \quad (7)$$

After obtaining the velocity-time equation, if its first derivative is taken, this value becomes equal to the acceleration of motion.

$$\frac{dv(t)}{dt} = a \quad (8)$$

In the 3D universe, motion with constant acceleration can be analyzed with the same principles in all three dimensions. In case the movement is in the vertical direction and no force other than the weight force acts on the object, the object moves with a constant acceleration with the acceleration of gravity. When the object is thrown upwards from the ground with a certain speed in the exact vertical direction, since the acceleration of gravity and the velocity vector will be in opposite directions, it makes a slowing motion first and its speed becomes zero at a certain height.

Then it starts accelerating downwards in the direction of gravitational acceleration from the point where the velocity is zero. During its entire motion, the kinematic equations can be used by taking h for displacement and g for acceleration. The time elapsed until the speed is zero is called the output time and is calculated with

$$t = \frac{v_0}{g} \quad (9)$$

In order to calculate the maximum height value, considering the fact that the final velocity will be 0 in equation 7 and the peak will be at this moment, the equation is revised and the maximum height equation becomes

$$h_{max} = \frac{v_0^2}{2g} \quad (10)$$

## Experimental Setup

Wooden block, Arduino UNO and HCSR-04 distance sensor were used in the study. The materials used are shown in the Figure 1.

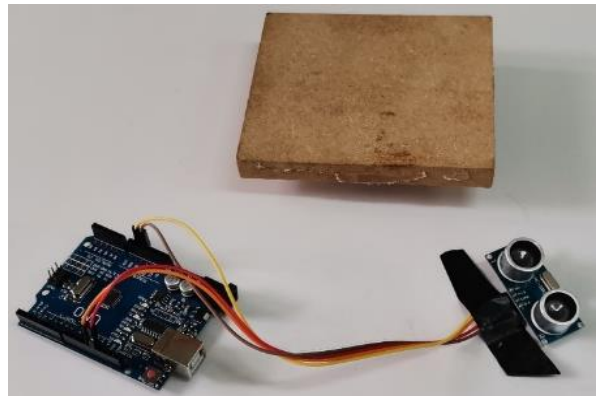

**Figure 1.** Materials

The connection between the sensor and the arduino is as in the Figure 2.

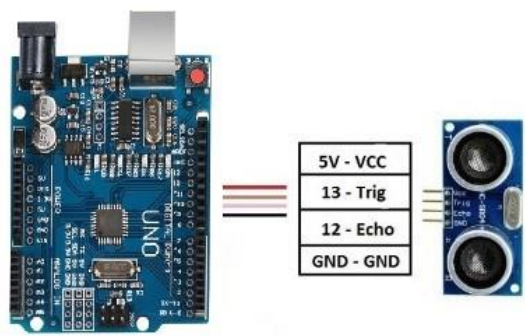

**Figure 2.** Connection diagram

The code used in programming is written in a way to print the distance values of the block standing on the vertical distance sensor, together with the time values, to the screen. This code is given as an attachment.

## Data analysis

In the experimental process, while the wooden block was standing at a certain height on the distance sensor, it was pushed upwards to gain speed and then released. In order to carry out detailed analyzes of kinematic equations and graphics, the data collected through the experimental process were divided into three and analyzed. In the Figure 3, the experimental scheme and three parts of the experimental process are shown in different colours.

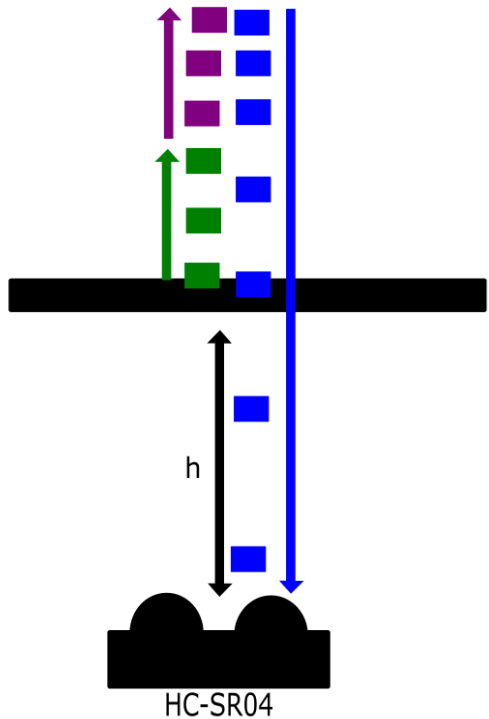

**Figure 3.** Diagram of the experimental process

The first part, shown in green, is the part about accelerating the wooden block by pushing it from the bottom up. From the moment it was released after being accelerated, the wooden block made a shooting motion from the bottom up. If the bottom-up shooting motion is also examined in two parts, the second part of the experimental process, shown in purple, is upward deceleration, and the third part, shown in blue, is free fall. The time-dependent graph of the height drawn using the data obtained during the whole process is as in the Figure 4.

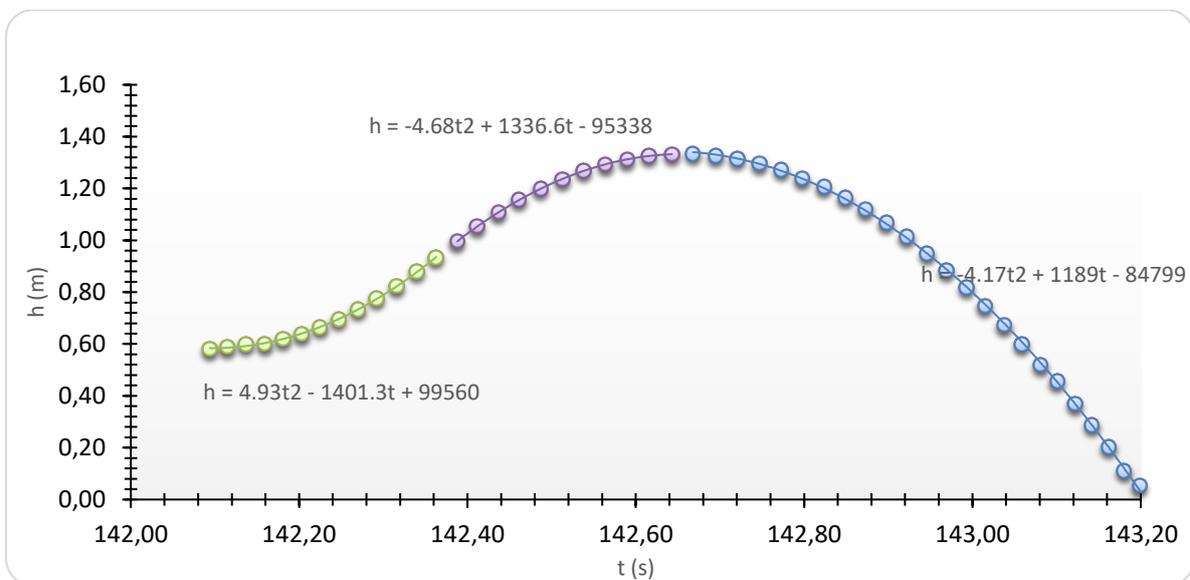

**Figure 4.** Time-dependent graph of the height of the wooden block above the distance sensor during the experimental process

First of all, three different position and time value pairs were selected on the graph in order to make applications related to kinematic concepts. One of these values should be the values when the maximum height is reached. Using the values in Figure 4, the displacement between the first point and the last point is determined $\overrightarrow{\Delta h} = 0.37 - 0.93 = -0.56$ m downwards, average speed determined as $\overrightarrow{\Delta v} = \frac{-0.56}{0.76} = -0.74$ m/s downward, distance traveled calculated as $x = (1.33 - 0.93) + (1.33 - 0.37) = 1.36$ m and average speed found as $v = \frac{1.26}{0.74} = 1.79$ m/s.

The equations of variation of the positions of all three sections with respect to time are shown in the Figure 4. By taking the first derivatives of these equations with respect to time, the time-dependent variation of velocity equations were found, and the instantaneous velocities were determined by substituting the time values in these equations. The graph of the change of velocity by time, drawn using the determined instantaneous speeds, is as in the Figure 5.

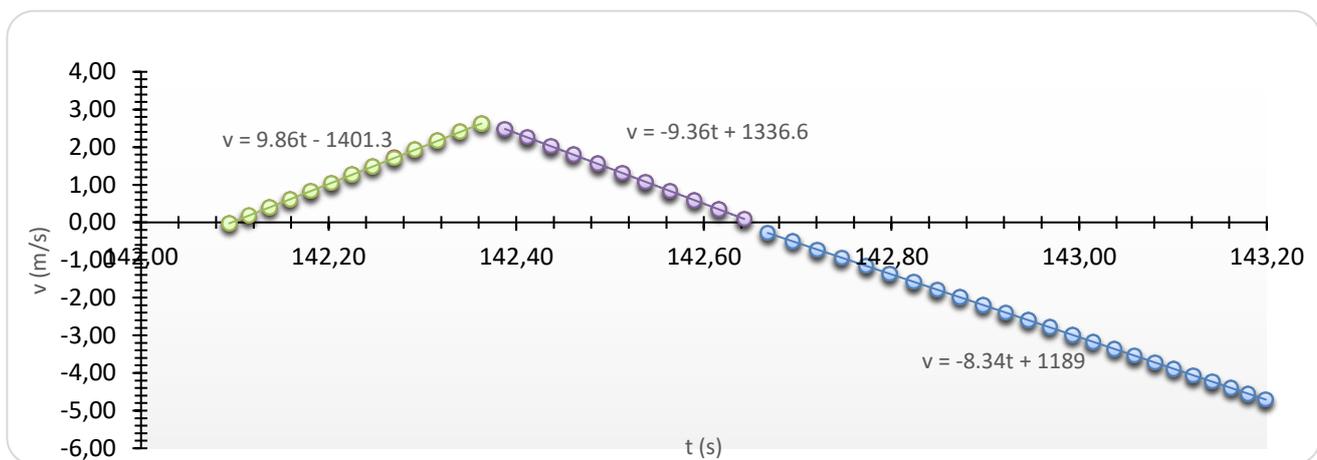

**Figure 5.** Time-dependent graph of the velocity of the wooden block during the entire movement

The first derivative of the equations of variation of velocity with respect to time in Figure 5 gives the acceleration. From here, the acceleration of the first, second and third parts of the movement found as +9.86 m/s², -9.36 m/s² and -8.34 m/s² respectively. The net force acting on the wooden at second and third part is the weight force and its acceleration is theoretically equal to the gravitational acceleration -9.81 m/s². It is seen that the acceleration values found at the second and third parts are close to the gravitational acceleration with an acceptable margin of error.

When the wooden block starts to move from the bottom up, the speed magnitude is 2.63 m/s and the magnitude of the deceleration acceleration is -9.36 m/s². Using these values, the time taken to reach the maximum height and the maximum height were calculated as 0.27 s and 0.35 m by using equation 9 and 10. The values determined using the collected data are 142.64-142.36= 0.28 s and 1.33-0.93=0.40 m. Both results showed high agreement with the results found by calculations. In order to analyse the kinematic equations, the position and velocity values at these start and end moments will be used in the motion between 142.64 s and 143.12 s. By taking the height as 1.33 m, initial velocity as 0.09 m/s, acceleration as -8.34 m/s² and elapsed time as 0.48 s, final velocity is calculated as -3.91 m/s using equation 6 and displacement calculated as -0.92 m using equation 5. When the figures are examined, it is seen that these values are -4.06 m/s and -0.96 m, respectively.

The movements of the wooden block while rising in the second section and falling in the third section are identical, and the instantaneous velocity magnitudes are the same and their directions are opposite when the height of the object is the same in these sections. This can be a situation that students may have difficulty in understanding during lectures in classes. For this reason, data analysis was performed on this situation in the study, and the velocity and height values obtained in the section where the object slowed down after being released and then accelerated downwards were used to reach the position-dependent change graph of the velocity given in the Figure 6.

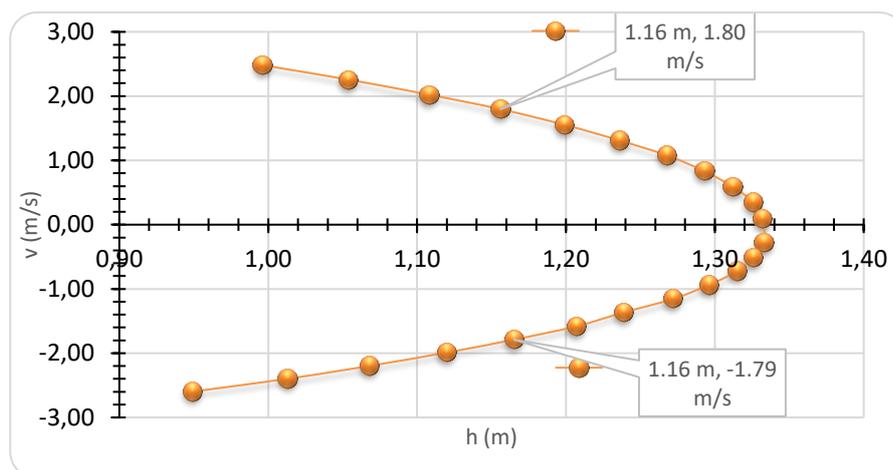

**Figure 6.** Positional graph of velocity during the throw straight up movement

When the graph in the figure is examined, it is clearly seen that the velocities at the same height are of the same magnitude and in opposite directions.

## Results

In the study, an experimental process including basic concepts analysis, equation analysis, and graphic analysis studies on 1-dimensional motion was designed. During these analyses, electronic connections, 3D material design, programming, data collection, and data analysis applications are carried out. With this feature, it has important benefits for all STEM fields. All the results obtained in the study are in very high agreement with the theoretical results. With the realization of such an experimental study with students in the classroom, more permanent and applicable gains can be obtained in the field of Physics through experimental applications. It may also be very important in terms of developing future inventors of technology. Determining the variables, such as instantaneous velocity and acceleration, can pave the way for very important activities. Experimental activities can be easily carried out with applications similar to those carried out in this study on issues such as energy conservation, Newton's laws analysis, and momentum. Variables such as instantaneous velocity and acceleration can also be determined with video analysis programs or expensive experimental tools. However, depending on the socioeconomic levels of the schools, there may be limitations in the implementation of such practices. The materials used in the study are easy to obtain and cost-effective. Taking these things into account, this kind of applications will help make sure that everyone has the same chances in school and will help students learn skills for the 21st century.

**Appendix: Code**

```
#define trigpin 13
#define echopin 12
void setup() {
Serial.begin(9600);
pinMode(trigpin, OUTPUT);
pinMode(echopin, INPUT);
}
void loop() {
digitalWrite(trigpin, LOW);
delay(5);
digitalWrite(trigpin, HIGH);
delay(10);
digitalWrite(trigpin, LOW);
float zaman = pulseIn (echopin, HIGH);
float x = (343.00*(zaman/1000000.00))/2;
float t = millis()/1000.00;
Serial.print(t,3);
Serial.print("*");
Serial.println(x,3);
}
```